\documentclass[reprint]{revtex4-1}
\usepackage[utf8]{inputenc}
\usepackage{hyperref}
\usepackage{graphicx}
\usepackage{physics}
\usepackage{float}
\usepackage{subfig}
\usepackage{subfiles}
\usepackage[normalem]{ulem}
\usepackage{yquant,braket}
\usepackage{todonotes}
\usepackage{multirow}

\usepackage{algorithm}
\usepackage{algorithmic}

\usepackage{enumitem,amssymb}
\newlist{todolist}{itemize}{2}
\setlist[todolist]{label=$\square$}
\usepackage{pifont}
\begin{document}

\title{Reducing runtime and error in VQE using deeper and noisier quantum circuits
}


\date{\today}

\author{Amara Katabarwa$^{1}$}
\author{Alex Kunitsa$^{1}$}
\author{Borja Peropadre$^{1}$}
\email{Current affiliation: IBM Yorktown Heights}
\author{Peter Johnson$^{1,}$}
\email{peter@zapatacomputing.com}

\affiliation{$^1$Zapata Computing, Boston, MA 02110 USA}


\begin{abstract}
The rapid development of noisy intermediate-scale quantum (NISQ) devices has raised the question of whether or not these devices will find commercial use.
Unfortunately, a major shortcoming of many proposed NISQ-amenable algorithms, such as the variational quantum eigensolver (VQE), is coming into view: the algorithms require too many independent quantum measurements to solve practical problems in a reasonable amount of time.
This motivates the central question of our work:
how might we speed up such algorithms in spite of the impact of error on NISQ computations?
We demonstrate on quantum hardware
that the estimation of expectation values, a core subroutine of many quantum algorithms including VQE, can be improved in terms of precision and accuracy by using a technique we call \emph{robust amplitude estimation}.
Consequently, this method reduces the runtime to achieve the same mean-squared error compared to the standard prepare-and-measure estimation method. 
The surprising result is that by using deeper, and therefore more error-prone, quantum circuits, we realize more accurate quantum computations in less time.
As the quality of quantum devices improves, this method will provide a proportional reduction in estimation runtime.
This technique may be used to speed up quantum computations into the regime of early fault-tolerant quantum computation and aid in the realization of quantum advantage.

\end{abstract}

\maketitle




Advances in quantum devices have fueled the development of near-term quantum algorithms \cite{o2016scalable, arute2019quantum, nam2020ground, rudolph2020generation}.
A driving question in the field is: can a quantum device achieve quantum advantage using such algorithms well-before the era of fault tolerance?
The variational quantum eigensolver (VQE), used for simulating chemistry and materials, has been a promising contender \cite{peruzzo2014variational}.
However, recent work has shown that the standard approach to implementing VQE requires exceedingly long runtimes for problem instances of industrial relevance \cite{gonthier2020identifying}.
This bottleneck is referred to as the ``measurement problem''.
Beyond quantum simulation, the measurement problem plagues
quantum machine learning and near-term quantum algorithms for linear algebra 
\cite{benedetti2019generative, somma2021complexity}.
Approaches to mitigating this measurement problem include Pauli-term grouping \cite{Izmaylov2020, zhao2020measurement, huggins2021efficient, crawford2021efficient, ralli2021implementation}
and improved allocation of measurements during parameterized circuit optimization \cite{sweke2020stochastic}.
While these techniques are helpful, the results of \cite{gonthier2020identifying} suggest that they are insufficient for resolving the measurement problem. 
A more scalable approach is to use quantum algorithms which reduce the runtime of statistical estimation.
Recent work \cite{wang2021minimizing} has developed a set of methods for \emph{robust amplitude estimation}: the task of estimating amplitudes in the presence of device error.
Such methods provide an interpolation between the performance of standard VQE and the performance of VQE in the fault-tolerant regime where quantum amplitude estimation is used.
The concept of such an interpolation was introduced in the context of VQE \cite{wang2019accelerated} and more recently has been explored in the context of quantum algorithms for finance \cite{yohichisuzukishumpeiunorudyraymondtomokitanakatamiyaonodera&naokiyamamoto2020, alcazar2021quantum, giurgica2021low}.

An outstanding question from the above theoretical work is: what are the limitations of robust amplitude estimation when implemented with real quantum devices?
In this paper we demonstrate that robust amplitude estimation already improves over standard VQE in the efficiency of estimating Pauli expectation values.
Furthermore, we find that robust amplitude estimation yields a degree of error mitigation without requiring additional samples (as other error mitigation methods do).
The most striking feature of this method is that it suggests the use of quite deep quantum circuits.
While in many settings one may want to keep overall circuit error rates below 10$\%$ or even $1\%$, 
our work suggests that runtime is minimized when using circuits with an overall circuit error rate of around $40\%$.
Through experiments using two qubits, we show that robust amplitude estimation yields a factor of four reduction in RMSE compared to standard sampling (as typically used in VQE), when spending the same allotted runtime.
In addition to a reduction in RMSE, we also find an increase in accuracy through a ten-fold bias reduction, demonstrating the error-mitigating capabilities of this method.


Robust amplitude estimation uses enhanced sampling to speed up the estimation of expectation values on noisy quantum computers. A detailed description of enhanced sampling techniques can be found in the following reference \cite{wang2021minimizing}.
In contrast to the standard sampling method used, for example, in VQE, enhanced sampling methods enable a reduction of estimation runtime proportional to improvements in the quality of the quantum hardware.
We explain the simplest variant of these techniques, which is used in our experiments.

Many quantum algorithms require the estimation of quantities encoded as expectation values
\begin{align}
\Pi = \bra{A} P \ket{A},
\label{eq:defPi}
\end{align}
where $\ket{A}=A\ket{0^n}$ in which $A$ is the ansatz circuit, and $P$ is an $n$-qubit Hermitian operator with eigenvalues $\pm 1$.
A straightforward estimation scheme entails repeatedly preparing state $\ket{A}$ and measuring $P$ (in the case of a Pauli operator, each qubit is measured in the corresponding Pauli basis and the parity of the outcomes gives the measured eigenvalue of $P$).
The estimate is taken to be the sample mean of the $\pm1$-valued outcomes.
We refer to this strategy as \emph{standard sampling} and note that this is typically used in VQE and other near-term quantum algorithms.

\begin{figure}[!ht]
\center
\includegraphics[width=9cm]{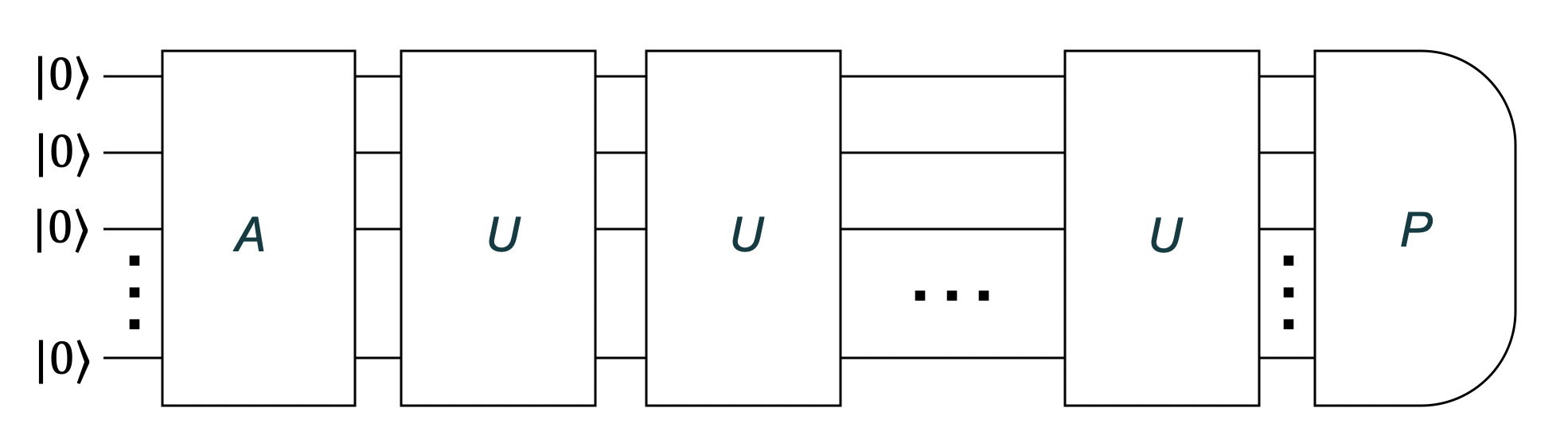}
\caption{This figure depicts the operations for generating measurement outcomes used in enhanced sampling. 
$A$ is the state preparation circuit, $P$ is the observable of interest, and $U$ is the Grover iterate comprised of $AR_0A^{\dagger}P$, with $R_0$ the reflection about the state $\ket{0}^{\otimes N}$.
} 
\label{fig:rae_circuit_diagram}
\end{figure}

The essence of enhanced sampling is to gather data from quantum states whose expectation values are functions of the expectation value of interest.
It modifies the standard sampling approach by using the following steps to generate measurement outcomes:
prepare the ansatz state $\ket{A}=A\ket{0^n}$, apply 
$L$ Grover layers $U=A(2\ket{0}\! \bra{0}^{\otimes n} - \mathbb{I} )A^{\dagger}P$, and then measure the Pauli observable $P$.
This sequence is depicted in Figure~\ref{fig:rae_circuit_diagram}.
Compared to standard sampling, the only substantially new operation introduced is the reflection about the initial state $R_0=2\ket{0}\! \bra{0}^{\otimes n} - \mathbb{I}$.

We estimate $\Pi=\langle A|P|A\rangle$ from a set of such measurement outcomes with varying $L$.
The estimate can be inferred from this data using a maximum likelihood estimation process (see discussion around Eq. \ref{eq:bayes}) based on the likelihood functions
\begin{align}
\label{eq:noiselesslf}
\mathbb{P}(\pm 1|\Pi;L)=\dfrac {1}{2}\left( 
1\pm T_{2L+1}(\Pi) \right),
\end{align}
where $T_{m}(x)=\cos(m\arccos(x))$ is the $m$-degree Chebyshev polynomial.
Note that the expression for $\mathbb{P}(- 1|\Pi;L)$ coincides with the Grover algorithm success probability due to the use of Grover layers $U$.
Enhanced sampling reduces the runtime needed to obtain an accurate estimate by drawing measurement outcomes whose likelihoods depend sensitively on the parameter of interest.

In practice, quantum computation is subject to errors.
These errors are caused by several sources, including decoherence to the thermal environment and limitations on the calibration of quantum gates.
We can estimate expectation values despite such error by incorporating a model of the effect of error on the likelihood function into the inference process.
If the model is sufficiently accurate and if we learn the model parameters to within a certain tolerance, then we may still be able to estimate the parameter of interest accurately.
We adopt the exponential decay model described in \cite{wang2021minimizing},
\begin{align}
\label{eq:decayed_likelihood}
\mathbb{P}(\pm1|\Pi,\lambda;L)=&\dfrac {1}{2}\left( 
1\pm e^{-(L+1/2)\lambda}T_{2L+1}(\Pi)\right),
\end{align}
where $\lambda$ is a decay parameter roughly proportional to the error rate of the elementary gates on the quantum device and, again, $T_{m}(x)=\cos(m\arccos(x))$ is the $m$-degree Chebyshev polynomial.
We emphasize that the model proposed here is an approximation to the actual relationship between the outcome likelihoods and the parameter of interest.
Therefore, an important question addressed by this work is: how well does robust amplitude estimation perform in practice when using this approximate model of the likelihood function?

In each step of the inference process we update a prior distribution $p(\Pi,\lambda)$, according to the parity of a bitstring outcome $s=\pm1$ generated from an $L$-layer enhanced sampling circuit and measurement, to obtain a posterior distribution
\begin{align}
\label{eq:bayes}
    p(\Pi,\lambda|s)=\frac{\mathbb{P}(s|\Pi,\lambda;L)p(\Pi,\lambda)}{\int d\Pi d\lambda \mathbb{P}(s|\Pi,\lambda;L)p(\Pi,\lambda)}.
\end{align}
After many measurements and Bayesian updates we determine the values $\hat{\Pi}$ and $\hat{\lambda}$ which maximize the posterior distribution and use this $\hat{\Pi}$ as the estimate of the parameter of interest.
The decay parameter $\lambda$ is treated as a nuisance parameter \cite{royall2000probability}, used as an intermediary for arriving at the maximum likelihood estimate (MLE) of $\Pi$.
Enhanced sampling methods vary in the means by which $L$ is chosen for each step and the way in which the Bayesian inference process is carried out numerically \cite{wang2021minimizing, yohichisuzukishumpeiunorudyraymondtomokitanakatamiyaonodera&naokiyamamoto2020, ericg.brownoktaygoktasw.k.tham2020, giurgica2020low}.

\begin{figure}[!ht]
\begin{tikzpicture}
    \begin{yquant}
qubit {$\ket{\reg_{\idx}}$} q[2];

box {$R_y(\frac{\pi}{2})$} q[0];
box {$R_x(\pi)$} q[1];
box {$R_x(-\frac{\pi}{2})$} q[1];
cnot q[1] | q[0];
box {$R_z(\theta)$} q[1];
cnot q[1] | q[0];
box {$R_y(-\frac{\pi}{2})$} q[0];
box {$R_x(\frac{\pi}{2})$} q[1];
measure q[0-1];

    \end{yquant}
\end{tikzpicture}
\caption{Circuit diagram of the VQE ansatz for molecular hydrogen we chose to use in our experiments. We chose the the circuit parameter to be $\theta=-6.0575$, which yields the ground state with respect to the molecular hydrogen Hamiltonian at 0.2 Angstroms.
The expectation values of $X_0X_1$ and $Y_0Y_1$ are both -0.2238 for this state.
}
 \label{fig:circuit_diagram}
\end{figure}
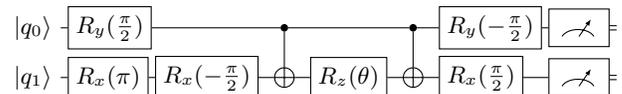

To demonstrate value of the robust amplitude estimation method in a setting of interest, we selected the ansatz circuit of \cite{PhysRevX.6.031007}, which was used to calculate the ground state energy of the hydrogen molecule. The circuit is reproduced in Figure~\ref{fig:circuit_diagram}.
For observables $P$, we restrict ourselves to Pauli strings $X_0X_1$, $Y_0Y_1$, which appear in the two-qubit hydrogen molecule Hamiltonian \cite{PhysRevX.6.031007}.

For each Pauli string, we generate parity measurement outcomes from the circuit of Figure \ref{fig:rae_circuit_diagram} for a range of $L$, the number of Grover iterates.
To generate the expectation value estimates, we process the measurement outcome data using Bayesian inference.
We carry out inference from the sampled data using numerical updates and maximum likelihood estimation (MLE) similar to \cite{yohichisuzukishumpeiunorudyraymondtomokitanakatamiyaonodera&naokiyamamoto2020, ericg.brownoktaygoktasw.k.tham2020}. We implement MLE by keeping track of a discretization of the posterior distributions resulting from Eq. \ref{eq:bayes}. 
After the data has been processed through these inference steps, we find a pair $(\lambda, \Pi )$ that maximizes the posterior distribution. We describe this procedure in more detail.
We define enhanced sampling circuits $\mathcal{C}=(\mathcal{C}_0, \mathcal{C}_1, \mathcal{C}_2, \dots \mathcal{C}_{L_{max}})$ and sets of parity outcomes $\mathcal{M}= (\mathcal{M}_{0}, \mathcal{M}_{1} \dots \mathcal{M}_{L_{max}})$, where $\mathcal{C}_i$ is a circuit of the type in Figure~\ref{fig:rae_circuit_diagram} with $i$ layers and $\mathcal{M}_i$ are the corresponding set of $\pm1$ parity outcomes resulting from $N_i$ total measurements and we set $N=\sum_{i=0}^{L_{max}} N_i$.
The algorithm for processing this outcome data into estimates is as follows:
\begin{algorithm}[H]
\label{alg:enhanced_sampling}
\begin{algorithmic}[1]
\algsetup{indent=2em}
\renewcommand{\thealgorithm}{} 
\floatname{algorithm}{Maximum Likelihood Estimation}
\caption{Estimate the value of $\Pi=\bra{A}P\ket{A}$ from a grid of candidate points
}
\label{ESC}

\REQUIRE $ \mathcal{M}= (\mathcal{M}_0, \mathcal{M}_1, \dots \mathcal{M}_{L_{max}})$  and prior distribution on a discrete grid $\mathcal{D}:(p,q)\rightarrow \mathbb{R}$
\ENSURE $ (\Pi, \lambda) $
\STATE Randomly sample, with duplicates, $M \leq N $ parity outcomes from $\mathcal{M}$ to get $ (m_0, m_1, \dots m_M)$ to create an ordered set $\mathcal{R}$ and a corresponding ordered set of $ \hat{\mathcal{T}} = (l_0, l_1,  \dots l_M)$ consisting of layer labels 
\FOR{$i=0$ to $i=M$}
    \IF{$m_i$=1}
      \item  $\mathcal{D}_{p,q}$  $\leftarrow \mathcal{D}_{p,q} \times   \frac{1}{2}(1 + e^{-(l_i+1/2 )\lambda_{p}}T_{2l_i +1}(\Pi_q))$
    \ELSE
    \item $\mathcal{D}_{p,q}$  $\leftarrow \mathcal{D}_{p,q} \times  \frac{1}{2}(1 - e^{-(l_i+1/2 )\lambda_p}T_{2l_i +1}(\Pi_q))$
    \ENDIF
\ENDFOR

\STATE Find $(\Pi_q^*, \lambda_p^*)$ such that $\mathcal{D}_{p,q}$ is  maximized over all $(p,q)$

\STATE $(\Pi, \lambda ) \leftarrow (\Pi_q^*, \lambda_p^*) $ 
\RETURN $(\Pi, \lambda )$
\end{algorithmic}
\end{algorithm}

In our experiments we chose to use 
two different IBM devices: $\textit{ibmq\_belem}$ and $\textit{ibmq\_manila}$.
We varied the number of layers between 0 and 10.
We generated a total of $N=90112$ samples, with $N_L=8192$ samples taken for each of the 11 layer numbers $L$.

We post-processed this data according to the maximum likelihood estimation algorithm shown above with $M=1000$ samples per estimate.
We run the maximum likelihood estimation algorithm 32 times, each time choosing a random sample of outcome data.
We choose 32 trials because the number of independent samples is 32 times the number of samples used per trial, making each trial nearly independent.

\begin{figure}[!ht]
    \centering
   \subfloat[The two plots compare the estimates of $\langle Y_0Y_1 \rangle $ and $\langle X_0X_1 \rangle $,  for different values of $L_{max}$. In the plots, each point is the mean among the 1000 estimates while the error bars indicate the standard deviation among these estimates. The dashed line represents the exact value of $\langle Y_0Y_1 \rangle$ and $\langle X_0X_1 \rangle$.]{ \includegraphics[width=\columnwidth,trim={.2cm 0 .5cm 0},clip]{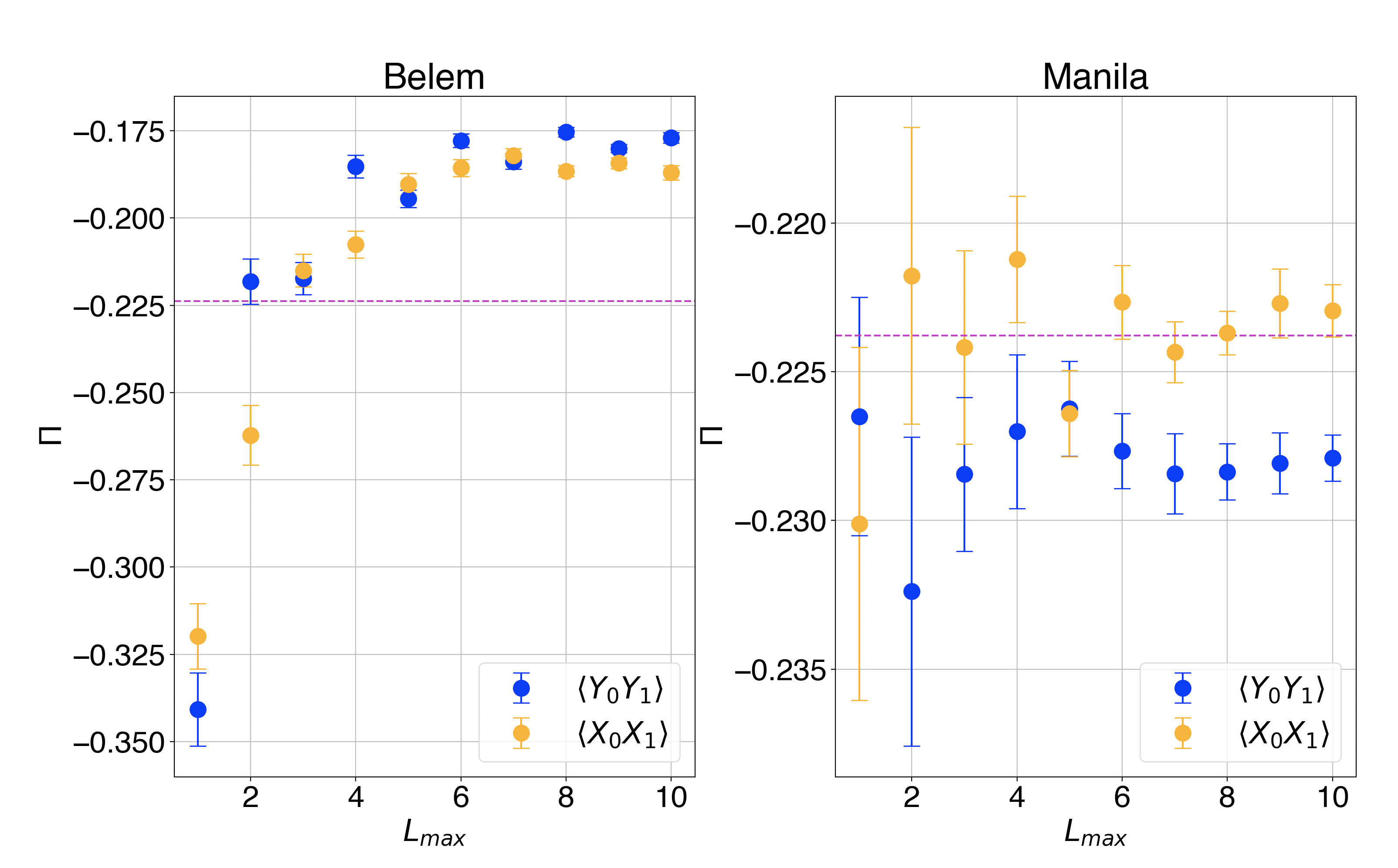}} \\
    \subfloat[In these plots we study the change in the RMSE of the estimates while varying $L_{max}$. For \textit{ibmq\_belem} devices, we see that the RMSE improves substantially until about $L_{max} \approx 3 $ after which we see now substantial gain in sampling power on the other hand \textit{ibmq\_manila} sees gains until $L_{max} \approx 5$.
    The error bars are computed as the square root of the standard deviation of the squared error from each trial, divided by the square root of the number of trials (i.e. 32).]{ \includegraphics[width=.9\columnwidth, trim={.5cm 0 2cm 0}]{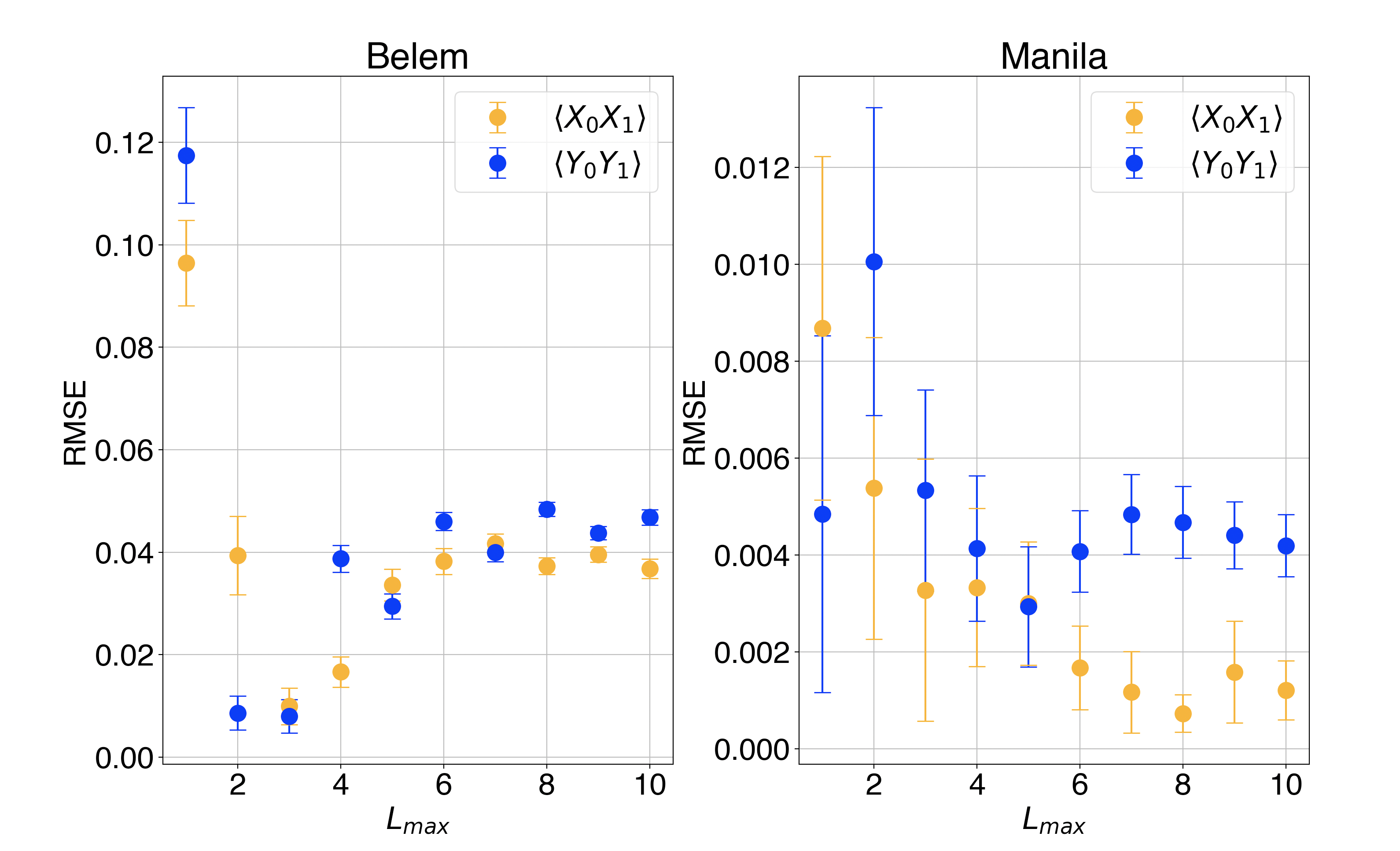}} \\
    \caption{Performance of robust amplitude estimation for estimating $\Pi$ for various $L_{max}$, the maximum number of Grover iterates used per circuit. Note that we do not include $L_{max}=0$ because we are unable to resolve two unknowns parameters (i.e. $\Pi$ and $\lambda$) with a single parameter frequency estimate from $L=0$ samples.}
    \label{fig:athens_santiago_5_layers}
\end{figure}

Figure \ref{fig:athens_santiago_5_layers} shows the estimated expectation value of $\langle Y_0Y_1 \rangle$ and $\langle X_0X_1 \rangle$  as we increase $L_{max}$, the maximum number of Grover iterates used. 
We observe a reduction in the RMSE as we increase the layers up to $L_{max}=3$ for \textit{ibmq\_belem}, beyond which there is no substantial reduction in RMSE, while for \textit{ibmq\_manila} the  improvement is observed
until $L_{max} \approx 5$.
We expect that in the regime of $L_{max}>3$ for \textit{ibmq\_belem} and $L_{max}>5$ for \textit{ibmq\_manila}, the deep circuits used have accrued too much error to yield valuable information.

\begin{figure}[!ht]
    \centering
    \includegraphics[width=1\columnwidth,trim={1cm 0 3cm 0}]{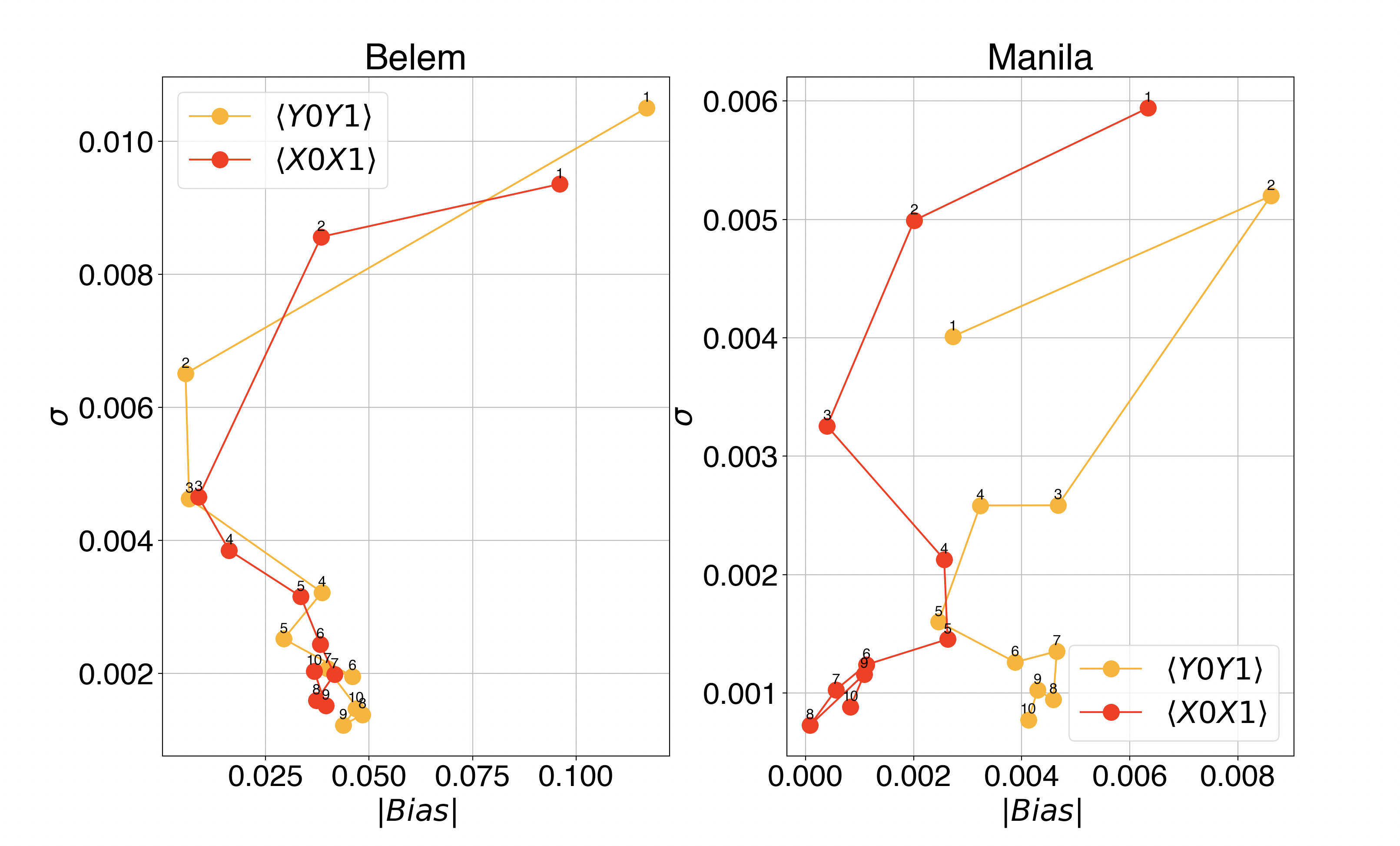}
    \caption{We show the change in bias and standard deviation, $\sigma$,  in the $\langle Y_0Y_1 \rangle$ and $\langle X_0X_1 \rangle$ expectation value estimates as the maximum number of Grover iterates is increased from 1 to 10.
    The RMSE is the distance from the origin.
    This plot shows that, in addition to an improvement in precision, RAE yields a reduction in bias indicating a degree of error mitigation.
    } 
    \label{fig:athens_santiago_precision_bias} 
\end{figure}

 In Figure \ref{fig:athens_santiago_precision_bias}  we demonstrate the surprising error mitigating capability of robust amplitude estimation. 
For each value of $L_{max}$, we plot the corresponding estimated bias and standard deviation of the expectation value estimates.
The substantial error mitigation is indicated by a nearly ten-fold reduction in the bias as the maximum layer number is increased from $L_{max}=1$.
We expect that the use of varying layer number in inferring the expectation value helps to average out the contributions from coherent error that differs from layer to layer.
 
These results lead to two main observations:
\begin{enumerate}[label= (\roman*)]
    \item Robust amplitude estimation can yield estimates with lower RMSE than standard sampling.
    \item Robust amplitude estimation exhibits a hitherto unforeseen error mitigation capability.

\end{enumerate}

It is worth remarking that these two improvements are achieved in parallel; this contrasts with traditional error mitigation methods (e.g. zero-noise extrapolation \cite{giurgica2020digital} and readout error mitigation schemes  like \cite{barron2020measurement}), where reduction in error comes at the cost of increased samples and run-time without the reduction in precision.

The surprising conclusion is the following: to reduce runtimes and error in estimation subroutines (e.g. as used in VQE) with near-term quantum devices we should employ deeper and noisier quantum circuits. These deeper quantum circuits use quantum amplification to increase the statistical power of the measurement outcome data.
This counter-intuitive conclusion can be understood in the following manner: a crucial ingredient in the noisy implementation of our algorithm is the likelihood function (c.f. Eq. \ref{eq:decayed_likelihood}) which incorporates a model of how noise (captured by the nuisance parameter) impacts the likelihoods of outcome data. 
The key point is that as long as this model is not far from the true likelihood function, the inference procedure can still yield accurate estimates, where the noise merely slows the rate of information gain as opposed to biasing the estimates.
This we believe highlights the importance of noise-tailoring for the NISQ era.

\begin{figure}
    \centering
    \includegraphics[width=95mm, height=50mm]{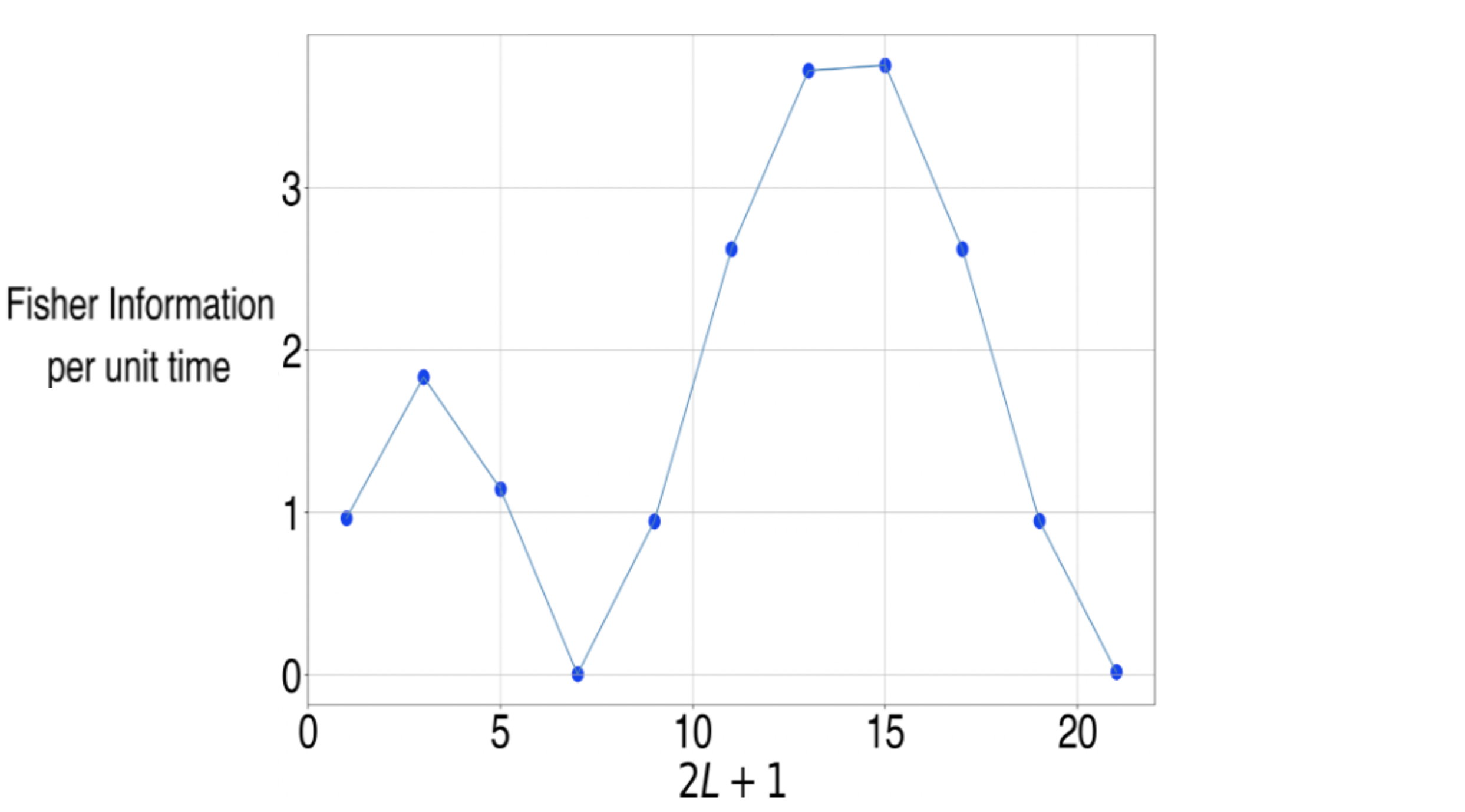}
    \caption{This plot shows the Fisher information per unit time for different numbers of Grover iterates. We use the true value of $\Pi=-0.22$ and $\lambda=0.08$, which are rough approximations to the values obtained through Bayesian inference. The maximum Fisher information rate is achieved for $L=7$, which is close to satisfying the rule of thumb for setting the optimal layer number: $2L+1\rightarrow\frac{1}{\lambda}$ (c.f. \cite{wang2021minimizing}). 
    We note that the plot provides an example of a ``dead spot'' as introduced in \cite{wang2021minimizing}: the exponential decay factor, introduced to model error in the quantum circuit, causes certain Grover circuits to yield unfavorable Fisher information.
    }
    \label{fig:fisher_info_rate}
\end{figure}

We have shown that robust amplitude estimation may increase precision and accuracy. We now investigate whether 
robust amplitude estimation achieves a better RMSE than standard VQE estimation when both methods are given the same allotted runtime.
We will restrict this exploration to the more-performant of the two IBM devices, $\textit{ibmq\_manila}$.
Furthermore, we will only consider the case of estimating $\langle X_0X_1\rangle$ and not $\langle Y_0Y_1\rangle$ because, according to Fig. \ref{fig:athens_santiago_5_layers}, RAE performed better for this case.
We expect that noise tailoring methods could help to remove the bias in the RAE estimates of $\langle Y_0Y_1\rangle$, but we leave this to future investigation.
In the standard sampling setting, we generate estimates assuming that $\lambda=0$, corresponding to the standard approach used in VQE.
To fix the runtimes to be equal, we account for the fact that the circuit depths (and therefore circuit times) are different in the two settings.
While each circuit in the VQE setting uses just one query to the ansatz circuit, each enhanced sampling circuit makes $(2L+1)$ queries to the ansatz.
Furthermore, it makes $L$ uses of the phase oracle, which we take to be roughly half the depth of the ansatz (in this two-qubit setting). 
To increase the rate of information gain in RAE, we choose to only draw samples from circuits whose likelihood functions yield large Fisher information per time:
\begin{align}
    \label{eq:fisher_info}
    \frac{{\cal I}}{T} = \frac{1}{x}\frac{e^{-\lambda x}\sin^2(x\theta)}{1-e^{-\lambda x}\cos^2(x\theta)},
\end{align}
where $x=2L+1$ and $\theta = \arccos\Pi$.
Setting $\lambda = 0.08$ and $\cos\theta = -0.22$, which are coarse approximations to the values resulting from maximum likelihood estimation, the Fisher information per time peaks around $L=1$ and $L=7$ as shown in Figure \ref{fig:fisher_info_rate}.
We choose to infer $\Pi$ using samples from $L=1, 5, 6, 7$;
We choose the total number of shots such that the total circuit time is equal to that of 12875 shots with $L=0$ (i.e. the VQE setting).
By taking $L=1, 5, 6,$ and $7$, the corresponding depths in units of the ansatz depth are $3.5$, $13.5$, $16$, and $18.5$. Thus, the average circuit depth (and, therefore, circuit time) will be 12.875 times that of the ansatz. 
Thus, to achieve the same runtime between the settings of VQE and robust amplitude estimation ($L=1, 5, 6, 7$), we allow $M=12875$ and $M=1000$ samples, respectively.

The results are reported in Table \ref{table:time_advantage}; we find that robust amplitude estimation yields roughly a four-fold reduction in RMSE and two-fold improvement in precision while costing roughly the same runtime.
Furthermore, there is roughly a ten-fold improvement in the bias, indicating a substantial degree of error mitigation achieved by robust amplitude estimation.
We note that this run-time costing has ignored the circuit latency times, which, if accounted for, would lead to an even further separation in performance between the two approaches.

\begin{center}
\begin{table}
\begin{tabular}{|c | c | c | c | c| }
\hline
 Method & RMSE & $\sigma$ (Prec.) & Bias (Acc.) & Samples  \\ 
 \hline \hline
 $\text{VQE}$ & 0.025(2) & 0.011(1) & 0.022(2) &  12,875 \\
 \hline
 $\text{RAE}$  & 0.0045(6) & 0.0043(5) & 0.0012(8) & 1,000   \\
\hline
\end{tabular}
\caption{This table compares the performance of the variational quantum eigensolver (VQE) and robust amplitude estimation (RAE) when given the same allotted runtime. We observe a factor of four improvement in RMSE, a factor of two improvement in precision, and a factor of ten improvement in accuracy (as measured by the bias) when estimating $X_0X_1$ on $ibmq\_manila$. 
Here, each RMSE, $\sigma$, and bias are estimated using 19 independent trials for VQE and 32 independent trials for RAE.}
\label{table:time_advantage}
\end{table}
\end{center}

In this work we have explored an approach to solving the so-called ``measurement problem'', which plagues many near-term quantum algorithms, including VQE.
It is believed that achieving quantum advantage with near-term quantum computers will require solving this problem \cite{gonthier2020identifying}.
The measurement problem can be attributed to the low statistical power of typical prepare-and-measure (referred to as ``standard sampling'') estimation methods.
Fortunately, quantum computing enables methods to increase the statistical power of estimation tasks through \emph{quantum amplification}, providing a path to solving the measurement problem.

We have investigated how recently-introduced techniques \cite{wang2021minimizing} involving quantum amplification, which we refer to as \emph{robust amplitude estimation}, stand to improve the performance of estimating expectation values with imperfect quantum computers.
The primary open question from previous work was: how do such methods perform in practice?
This work may be seen as a step towards assessing robust amplitude estimation \emph{in practice} as a viable solution to the measurement problem.

Using several different publicly-available IBM devices, we implemented both robust amplitude estimation and estimation with standard sampling for estimating two-qubit Pauli expectation values.
We found that robust amplitude estimation gives a substantial improvement in two distinct ways: reduction in runtime and error mitigation.

We hope that our findings encourage practitioners of near-term quantum algorithms to begin adopting enhanced sampling and robust amplitude estimation as valuable tools for their quantum computations.
As the demand for using quantum devices increases, quantum compute time will become more costly.
The runtime savings afforded by robust amplitude estimation translate directly to a monetary cost savings.

This work leaves a number of directions open for future investigation.
It remains to develop a rigorous analysis and model for the performance of these techniques as the system size is scaled to larger qubit counts.
Improving over the algorithm performance model in \cite{wang2021minimizing}, such a model would be very valuable for carrying out resource estimations and assessing prospects for quantum advantage.
Another important direction is to develop methods for estimating multiple parameters in parallel, which is a feature of standard sampling as used in VQE.
In future work  we plan to incorporate noise-tailoring techniques, such as randomized compiling \cite{wallman2016noise}, to make the actual likelihood function closer to the likelihood function model, improving the performance of estimation.
With each of these directions in mind and with the promising findings on robust amplitude estimation, we hope this work will encourage researchers to develop methods that are necessary for realizing useful quantum computation.

\textbf{Acknowledgements:} The authors thank J\'er\^ome Gonthier, Artur Izmaylov, Jens Eisert, Yudong Cao, Micha\l{} Stech\l{}y, and Dax Koh for helpful feedback on the manuscript. Experiments were run using the Zapata Computing Platform Orquestra\texttrademark.

\bibliographystyle{unsrt}
\bibliography{bibliography}

\end{document}